# Correlated Anarchy
# in Overlapping Wireless Networks

Panayotis Mertikopoulos and Aris L. Moustakas, *Senior Member, IEEE*

*Abstract*—We investigate the behavior of a large number of selfish users that are able to switch dynamically between multiple wireless access-points (possibly belonging to different standards) by introducing an iterated non-cooperative game. Users start out completely uneducated and naïve but, by using a fixed set of strategies to process a broadcasted training signal, they quickly evolve and converge to an evolutionarily stable equilibrium. Then, in order to measure efficiency in this steady state, we adapt the notion of the *price of anarchy* to our setting and we obtain an explicit analytic estimate for it by using methods from statistical physics (namely the theory of replicas). Surprisingly, we find that the price of anarchy does not depend on the specifics of the wireless nodes (e.g. spectral efficiency) but only on the number of strategies per user and a particular combination of the number of nodes, the number of users and the size of the training signal. Finally, we map this game to the well-studied *minority game*, generalizing its analysis to an arbitrary number of choices.

*Index Terms*—Wireless networks, Nash equilibrium, correlated equilibrium, price of anarchy, evolutionary game, replicas

## I. INTRODUCTION

AS a result of the massive deployment of IEEE 802.11 wireless networks, and in the presence of large-scale mobile third-generation systems, mobile users often have several choices of overlapping networks to connect to. In fact, devices that support multiple standards already exist and, additionally, significant progress has been made towards creating flexible radio devices capable of connecting to *any* existing standard [1]. It is thus reasonable to expect that, in the near future, users will be able to switch dynamically between different networks.

In such a setting, even though users have several choices to connect to, they still have to compete against each other for the finite resources of the combined network. Hence, this situation can be modelled using non-cooperative game theory, a practice that is rapidly becoming one of the main tools in the analysis of wireless networks. For example, game-theoretic techniques were used to optimize transmission probabilities in [2] and to calculate the optimal power allocation [3]–[6] or the optimal transmitting carrier in [7]. The authors of [8] and [9] studied the possibility of connecting to several access points using a single WLAN card; the selfish behavior of service providers was analyzed in [10]–[12] and, recently, even the effects of pricing were examined in [13]–[16] using game theory.

P. Mertikopoulos and A. Moustakas are with the Physics Department, National and Kapodistrian University of Athens, 157 84 Athens, Greece.

This research was supported in part by the European Commission under grants EU-MIRG-CT-2005-030833 (PHYSCOM), EU-FET-FP6-IST-034413 (NetReFound) and EU-IST-2006-27960 (URANUS); the first author was also supported by the Empirikion Foundation of Athens, Greece. Part of this material was presented at the 2007 GameComm Workshop in Nantes, France.

The scenario that we consider is an unregulated network where a large number of $N$ *heterogeneous* users (e.g. mobile devices) connect wirelessly to one of $B$ nodes (perhaps with different standards). All users wish to maximize their individual downlink throughput but each has a different approach: e.g. users may have different tolerance for delay, or may wish to employ different "betting" schemes to download data at the lowest price. So, in general, users have different strategies, fixed at the outset of the game, and unknown to the rest.

Now, given the users' competition for the nodes' limited resources, it is not clear how they can reach an organized state in the absence of a central coordinating entity. One possible way to overcome this hurdle is if users base their decisions on a "training" signal, e.g. a random signal that is synchronously broadcasted by the nodes and received by all the users. Then, as this affair is iterated, one might hope that sophisticated users develop an insight into how other users respond to the same stimulus and, eventually, learn to coordinate their actions. This was precisely the seminal idea behind Aumann's work in [17]: players base their decisions on their observations of the "states of the world" and reach a *correlated equilibrium*.

Similar games have also been studied in econophysics, particularly after the introduction of the *El Farol* problem in [18] and the development of the *minority game* in [19]. In both these games, players "buy" or "sell" and are rewarded when they land in the minority. Again, the key idea is that in order to decide what to do, players record and process the game's history with the aid of some predetermined strategies. Then, by employing more often the strategies that perform better, they quickly converge to an equilibrium which (in an unexpected twist) turns out to be oblivious to the source of the players' observations [20]. In fact, it was shown in [21] that what matters is simply the *amount* of feedback that players receive and the number of strategies they use to process it.

As in [22], our scenario stands to gain a lot from such an approach. Hence, our main goal will be to expound this scheme in a way appropriate for selfish users in an unregulated wireless network. The first step towards this is to generalize and adapt the minority game of [21] to our setting: this is done in section II where we introduce the *Simplex Game*. Next, in section III, we characterize the game's equilibria and compare them to the socially optimal state. From this comparison emerges the game's *price of anarchy*, a notion first described in [23] and which measures the distance between anarchy (equilibria) and efficiency (optimal states).

Our first important result is obtained in section IV: by iterating the game based on the scheme of exponential learning, we find that *players converge to an evolutionarily stable equilibrium* (theorem 11). Then, having established convergence, we proceed in section V to harvest the game's price of anarchy. Quite unexpectedly, we find that *the price of anarchy is unaffected by disparities in the nodes' characteristics* (theorem 14). Moreover, we also derive an analytic expression for the price of anarchy based on the method of replicas from statistical physics. This allows us to study the effect of the various parameters on the network's performance, an analysis which we supplement with numerical experiments. As a byproduct, this generalizes the results of the traditional (binary) minority game to an arbitrary number of choices.

Some calculational details that would detract one's focus from the main discussion have been deferred to the appendices at the end. Finally, as far as notational conventions go, we will denote the standard $(n-1)$-dimensional simplex of $\mathbb{R}^n$ by $\Delta_n = \{ \mathbf{x} \in \mathbb{R}^n : x_i \geq 0 \text{ and } \sum_i x_i = 1 \}$; also, we will employ the game-theoretic shorthand: $(x_{-i}; y) = (x_1 \ldots y \ldots x_n)$.

## II. The Simplex Game

To model the scenario that we described in the introduction, we consider $N$ users that may choose one of $B$ nodes, each characterized by a single user spectral efficiency $c_r$. In this case, if $N_r$ users connect to node $r$, their throughput will be:

$$T_r = \frac{c_r}{N_r} \tag{1}$$

(for simplicity we assume that users have the same transmission characteristics).

Despite the simplicity of this throughput model, it has been shown to be of the correct form for TCP and UDP protocols in IEEE 802.11b systems, if we limit ourselves to a single class of users [9]. Furthermore, in the case of third-generation best-effort systems, the realistic total cell-service throughput is approximately constant beyond a certain number of connected users [24]. Thus, (1) is a reasonable approximation for the user throughput of single-class mobiles.

In fact, equation (1) is flexible enough to account for parameters that affect a user's bias towards a node; e.g. we can incorporate pricing by modifying $c_r$ to $c_r(1 - p_r)$ where $p_r$ reflects the price per bit. So, we may renormalize (1) to:

$$\hat{u}_r = \frac{y_r N}{N_r} \tag{2}$$

where the coefficients $y_r$ are normalized to unity ($\sum_{r=1}^B y_r = 1$) and represent the "effective strength" of node $r$ in terms of its attributes and characteristics. Clearly, nodes can modify this "strength" score, in order to maximize their gain; however, this is assumed to take place at slower time-scales and, hence, these strengths can be assumed to remain constant.[1]

We may now note that the core constituents of a congestion game are all present: $N$ *players* (users) are asked to choose one of $B$ *facilities* (nodes), their payoff given by the throughput (2). From this standpoint, the "fairest" user distribution is the Nash allocation of $y_r N$ users to node $r$: when distributed this way, users receive a payoff of $\hat{u}_0 = 1$ and no one could hope to earn more from a unilateral deviation (comparably to the

"water-filling" of e.g. [25]). As a result, the users' discomfort can be gauged by contrasting their payoff to the Nash value:

$$\hat{u}_r - \hat{u}_0 = \frac{y_r N}{y_r N + N_r - y_r N} - 1 = \frac{y_r N - N_r}{y_r N} + O(1/N) \tag{3}$$

So, if we focus on the leading term of (3) and introduce:

$$u_r = 1 - \frac{N_r}{y_r N} \tag{4}$$

we may easily that the Nash equilibria of the game remain invariant under this linearization. In other words, the payoffs (2) and (4) will be equivalent in terms of social fairness.[2]

Thanks to this linearization, we may express a user's payoff in a particularly revealing form. However, to accomplish this, we first need to introduce a collection of $B$ vectors in $\mathbb{R}^{B-1}$ with which to model the nodes:

*Definition 1:* Let $\mathbf{y} = (y_1 \ldots y_B) \in \text{Int}(\Delta_B)$ be a strength distribution for $B$ nodes.[3] A $\mathbf{y}$-*simplex* (or $\mathbf{y}$-*appropriate simplex*) is a collection $\mathscr{B} = \{ \mathbf{q}_r \}_{r=1}^B \subseteq \mathbb{R}^{B-1}$ such that, for $r, l = 1 \ldots B$:

$$\mathbf{q}_r \cdot \mathbf{q}_l = -1 + \frac{\delta_{rl}}{\sqrt{y_r y_l}} \tag{5}$$

Admittedly, this definition is rather opaque[4] but, fortunately, the geometric picture is much clearer:

*Lemma 2:* Let $\mathscr{B} = \{ \mathbf{q}_r \}_{r=1}^B$ be a $\mathbf{y}$-appropriate simplex for some $\mathbf{y} \in \text{Int}(\Delta_B)$. Then: $\sum_r y_r \mathbf{q}_r = 0$; also: $\sum_r y_r \mathbf{q}_r^2 = B - 1$.

*Proof:* To establish the first part, note that: $\left( \sum_{r=1}^B y_r \mathbf{q}_r \right)^2 = \sum_{r,l=1}^B y_r y_l \mathbf{q}_r \cdot \mathbf{q}_l = \sum_{r,l=1}^B y_r y_l \left( -1 + \delta_{rl} / \sqrt{y_r y_l} \right) = 0$. As for the second part, it is just a straightforward application of (5). ∎

In other words, a $\mathbf{y}$-simplex is just like a standard simplex with vertices "weighted" by the strengths $y_r$.[5] So, if $N_r$ players choose $\mathbf{q}_r$, we may consider the *aggregate bet* $\mathbf{q} = \sum_{l=1}^B N_l \mathbf{q}_l$ and obtain by (5): $\mathbf{q}_r \cdot \mathbf{q} = \sum_{l=1}^B N_l \mathbf{q}_r \cdot \mathbf{q}_l = -N \left( 1 - \frac{N_r}{y_r N} \right)$. We then get the very useful expression for the payoff (4):

$$u_r = 1 - \frac{N_r}{y_r N} = -\frac{1}{N} \mathbf{q}_r \cdot \mathbf{q} = -\frac{1}{N} \mathbf{q}_r \cdot \sum_{i=1}^N \mathbf{q}_{r_i} \tag{6}$$

where $r_i$ indicates the choice of player $i$. In this way, lemma 2 shows that Nash equilibria will be characterized by:

$$\mathbf{q} = \sum_{i=1}^N \mathbf{q}_{r_i} = \sum_{r=1}^B y_r N \, \mathbf{q}_r = 0 \tag{7}$$

i.e. *the game will be at equilibrium when the players' choices balance out the weights $y_r$ at the vertices of the simplex.*

Unfortunately, it remains unclear how this Nash allocation can be achieved in an unregulated system. For this reason, we will introduce a coordination mechanism akin to the one proposed by Aumann in his seminal paper [17]. In a nutshell, Aumann's scheme is that players observe the random events $\gamma$ that transpire in some sample space $\Gamma$ (the "states of the world") and then place their bets based on these observations. In other words, players' decisions are ordained by their (correlated) *strategies* $f_i$, i.e. functions on $\Gamma$ that convert events ("states") $\gamma \in \Gamma$ to *actions* (betting suggestions) $f_i(\gamma)$.

---

[1] Obviously, nodes of zero strength (e.g. negligible spectral efficiency) will not appeal to any reasonable user and can be dropped from the analysis.

[2] This is also verified by our numerical experiments (see figure 1).

[3] To clear up any confusion: $\text{Int}(\Delta_B) = \{ \mathbf{y} \in \mathbb{R}^B : y_r > 0 \text{ and } \sum_r y_r = 1 \}$.

[4] In fact, it is not even clear that the definition is not vacuous. This is shown in appendix A: $\mathbf{y}$-simplices are pretty easy to construct for any $\mathbf{y} \in \text{Int}(\Delta_B)$.

[5] One could also ask here why we insist that $\mathbf{y}$-simplices be embedded in $\mathbb{R}^{B-1}$ instead of $\mathbb{R}^B$. The reason for this is quite subtle and hinges on the fact that we need $\mathscr{B}$ to span the space it is embedded in, so that we may apply the Hubbard-Stratonovich transformation (see appendix B).

Inspired by [17] (and also [21]), we propose that a broadcast beacon transmit a *training signal* $m$, drawn from some (discrete) sample space $\mathcal{M}$. For example, the nodes could be synchronously broadcasting the same integer $m \in \{1 \ldots M\}$, drawn from a uniform random sequence that is arbitrated e.g. by a government agency such as the FCC in the US. To process this signal, user $i$ has at his disposal $S$ $\mathcal{B}$-valued random variables $\mathbf{c}_{is} : \mathcal{M} \to \mathcal{B}$ ($s = 1 \ldots S$):[6] these are the $i^{\text{th}}$ user's *strategies*, used to convert the signal $m$ to an *action* $\mathbf{c}_{is}(m) \equiv \mathbf{c}_{is}^m \in \mathcal{B}$. So, if user $i$ employs strategy $s_i$, the collection of maps $\{\mathbf{c}_{is_i} : \mathcal{M} \to \mathcal{B}\}_{i=1}^{N}$ will be a correlated strategy in the sense of [17] (contrast with $\{f_i\}_{i=1}^{N}$ above).

However, unlike [17], we cannot assume that users develop their strategies after careful contemplation on the "states of the world". After all, it is quite unlikely that a user will have much time to think in the fast-paced realm of wireless networks. Consequently, when the game begins, we envision that each user randomly "preprograms" $S$ strategies, drawn randomly from all the possible $B^M$ maps $\mathcal{M} \to \mathcal{B}$. Of course, since we assume users to be *heterogeneous*, they will program their strategies in wildly different ways and independently of one another. Still, rational users will exhibit a predisposition towards stronger nodes; to account for this, we will posit that:

$$\mathrm{P}(\mathbf{c}_{is}^m = \mathbf{q}_r) = y_r \qquad (8)$$

i.e. the probability that user $i$ *programs* node $\mathbf{q}_r$ as response to the signal $m$ is just the node's strength $y_r$. In effect, strategies are picked in *anticipation* of competition with other users: specifically, if each user were expecting to play alone, he would have picked strategies that lead to the strongest node.

We may now summarize the above in a formal definition:

*Definition 3:* Let $\mathbf{y} \in \mathrm{Int}(\Delta_B)$ be a strength distribution for $B$ nodes. Then, a **y**-*appropriate simplex game* $\mathfrak{G}$ consists of:

1) the set of *players*: $\mathcal{N} = \{1 \ldots N\}$;
2) the set of *nodes*: $\mathcal{B} = \{\mathbf{q}_r\}_{r=1}^{B}$, where $\mathcal{B}$ is a **y**-simplex;
3) the set of *signals*: $\mathcal{M} = \{1 \ldots M\}$, endowed with the uniform measure $\varrho_0(m) = \frac{1}{M}$; the ratio $\lambda = \frac{M}{N}$ will be called the *training parameter* of the game;
4) the set of *strategy choices*: $\mathcal{S} = \{1 \ldots S\}$; also, for each player $i \in \mathcal{N}$, a probability measure $p_i(s) \equiv p_{is}$ on $\mathcal{S}$ ($\sum_{s=1}^{S} p_{is} = 1$): these are the players' *mixed strategies*;
5) a *strategy matrix* $\mathbf{c} : \mathcal{N} \times \mathcal{S} \times \mathcal{M} \to \mathcal{B}$ where $\mathbf{c}(i, s, m) \equiv \mathbf{c}_{is}^m \in \mathcal{B}$ is the node that the $s^{\text{th}}$ strategy of user $i$ indicates as response to the signal $m \in \mathcal{M}$; the entries of $\mathbf{c}$ are drawn randomly based on: $\mathrm{P}(\mathbf{c}_{is}^m = \mathbf{q}_r) = y_r$.

Moreover, we endow $\Omega = \mathcal{M} \times \mathcal{S}^N$ with the product measure $\varrho_0 \times \prod_{i=1}^{N} p_i$ and define the following:

6) an *instance* of $\mathfrak{G}$ is an event $\omega = (m, s_1, \ldots s_N)$ of $\Omega$;
7) the *bet* of player $i$ is the $\mathcal{B}$-valued random variable: $\mathbf{b}_i(\omega) = \mathbf{c}(i, s_i, m)$; also, the *aggregate bet* is: $\mathbf{b} = \sum_{i=1}^{N} \mathbf{b}_i$;
8) the *payoff* for player $i$ is the r.v.: $u_i = -\frac{1}{N} \mathbf{b}_i \cdot \mathbf{b}$.

Thus, similarly to the minority game of [19] and [21], the sequence of events that we intuitively envision is:[7]

- in the "initialization" phase (steps 1-5), players program their strategies by drawing the strategy matrix $\mathbf{c}$;
- in step 6, the signal $m$ is broadcasted and, based on $p_i$, players pick a strategy $s \in \mathcal{S}$ to process it with: $p_{is}$ is the probability that user $i$ employs his $s^{\text{th}}$ strategy;
- in steps 7-8, players connect to the nodes that their strategies indicate ($\mathbf{b}_i(m, s_1 \ldots s_N) = \mathbf{c}_{is_i}^m$) and receive the linear payoff (4): the $N_r$ users that end up connecting to node $\mathbf{q}_r$ receives: $-\frac{1}{N} \mathbf{q}_r \cdot \sum_l N_l \mathbf{q}_l = -\frac{1}{N} \frac{N_r}{N}$;
- the game is iterated by repeating steps 6-8.

As usual, the payoff that corresponds to the (mixed) strategy profile $p = (p_1 \ldots p_N)$ will be the multilinear extension: $u_i(m, p) = \sum_{\{s\}} p_{1s_1} \ldots p_{Ns_N} u_i(m, s_1 \ldots s_N)$. To avoid carrying cumbersome sums like this, we will follow the notation of [21] and use $\langle \cdot \rangle$ to indicate expectations over a particular player's mixed strategy: $\langle v_i \rangle = \sum_s p_{is} v_{is}$; also, we will use an overline to denote averaging over the training signals, as in: $\overline{a} = \frac{1}{M} \sum_m a^m$.

## III. Selfishness and Efficiency

Clearly, the only way that selfish users who seek to maximize their individual throughput can come to an unmediated understanding is by reaching an equilibrial state that discourages unilateral deviation. But, since there is a palpable difference between the users' *strategic decisions* ($s \in \mathcal{S}$) and the *tactical actions* they take based on them ($\mathbf{c}_{is}^m \in \mathcal{B}$), one would naturally expect the situation to be somewhat involved.

### A. Notions of Equilibrium

Indeed, it should not come as a surprise that this dichotomy between strategies and actions is reflected on the game's equilibria. On the one hand, we have already encountered the game's tactical equilibrium: it corresponds to the Nash allocation of $y_r N$ users to node $r$. On the other hand, given that users only control their strategic choices, we should also examine Aumann's strategic notion of a *correlated equilibrium*.

To that end, recall that a correlated strategy is a collection $f = \{f_i\}_{i=1}^{N}$ of maps $f_i : \mathcal{M} \to \mathcal{B}$ (one for each player) that convert the signal $m$ to a betting suggestion $f_i(m) \in \mathcal{B}$. We will then say that a (pure) correlated strategy $f$ is at *equilibrium for player $i$* when, for all perturbations $(f_{-i}; g_i) = (f_1 \ldots g_i \ldots f_N)$ of $f$, player $i$ gains more (on average) by sticking to $f_i$, i.e. $u_i(f) \geq u_i(f_{-i}; g_i)$. When this is true for all players $i \in \mathcal{N}$, $f$ will be called a *correlated equilibrium*.

As we saw before, if user $i$ picks his $s_i^{\text{th}}$ strategy, the collection $\{\mathbf{c}_{is_i} : \mathcal{M} \to \mathcal{B}\}_{i=1}^{N}$ is a correlated strategy, but the converse need not hold: in general, not every correlated strategy can be recovered from the limited number of preprogrammed strategic choices.[8] Thus, users will no longer be able to consider *all* perturbations of a given strategy, and we are led to:

*Definition 4:* In the setting of definition 3, a strategy profile $p = (p_1 \ldots p_N)$ is a *constrained* correlated equilibrium when, for all strategy choices $s \in \mathcal{S}$ and for all players $i \in \mathcal{N}$:

$$\frac{1}{M} \sum_m u_i(m, p) \geq \frac{1}{M} \sum_m u_i(m, p_{-i}; s). \qquad (9)$$

---

[6]We are assuming that $S$ is the same for all users for the sake of simplicity.

[7]It is important to note here that, for 2 identical nodes ($\mathcal{B} = \{-1, 1\}$), the simplex game reduces exactly to the original minority game of [21].

[8]There is a total of $B^{MN}$ correlated strategies but users can recover at most $S^N$ of them. In fact, this is why preprogramming is so useful: it would be highly unreasonable to expect a given user to process in a timely fashion the exponentially growing number of $B^M$ (as compared to $S$) strategies.

The set of all such equilibria of $\mathfrak{G}$ will be denoted by $\Delta^0(\mathfrak{G})$.

In our setting, a (constrained) correlated equilibrium is what represents anarchy: with no one to manage the users' selfish desires, the only thing that deters them from unilateral deviation is their expectation of (average) loss. Conceptually, this is pretty similar to the notion of a Nash equilibrium, the main difference being that in a correlated equilibrium we are averaging the payoff over the training signals. This analogy will be very useful to us and we will make it precise by introducing the *correlated form* of the simplex game:

*Definition 5:* The *correlated form* of a simplex game $\mathfrak{G}$ is a game $\mathfrak{G}^*$ with the same set of players $\mathcal{N} = \{1 \ldots N\}$, each one choosing an action from $\mathcal{S} = \{1 \ldots S\}$ for a payoff of:

$$u_i^*(s_1 \ldots s_N) = \frac{1}{M} \sum_m u_i(m, s_1 \ldots s_N) \qquad (10)$$

In short, the payoff that players receive in the correlated game is their throughput averaged over a rotation of the training signals. Then, an important consequence of definition 4 is that *the constrained correlated equilibria of a simplex game $\mathfrak{G}$ are precisely the Nash equilibria of its correlated form $\mathfrak{G}^*$.*

### B. Harvesting the Equilibria

So, our next goal will be to understand the Nash equilibria of $\mathfrak{G}^*$. To begin with, a brief calculation shows that the payoff $u_i^*$ for a mixed profile $p = (p_1 \ldots p_N)$ is:

$$u_i^*(p_1 \ldots p_N) = -\frac{1}{N} \left\{ \langle \mathbf{c}_i \cdot \sum_{j \neq i} \langle \mathbf{c}_j \rangle \rangle + \overline{\langle \mathbf{c}_i^2 \rangle} \right\} \qquad (11)$$

(the averaging notations $\overline{(\cdot)}$ and $\langle \cdot \rangle$ being as in the end of section II). Thus, given the similarities of our game with congestion games, it might be hoped that its Nash equilibria can be harvested by means of a *potential function*, i.e. a function that measures the payoff difference between users' individual strategies [26]. More concretely, a potential $U$ satisfies: $u_i^*(p_{-i}; s_1) - u_i^*(p_{-i}; s_2) = U(p_{-i}; s_1) - U(p_{-i}; s_2)$ for any mixed profile $p = (p_1 \ldots p_S)$ and any two strategic choices $s_{1,2}$ of player $i$. Obviously then, if a potential function exists, its local maxima will be Nash equilibria of the game.

But, unfortunately, since $\mathfrak{G}^*$ does not have an exact congestion structure, it is not clear how to construct such a potential. Nevertheless, a good candidate is the game's aggregate payoff $u^* = \sum_i u_i^*$. In fact, if player $i$ chooses strategy $s$, $u^*$ becomes:

$$u^*(p_{-i}; s) = -\frac{1}{N} \left( \sum_{\substack{l,k \neq i \\ l \neq k}} \overline{\langle \mathbf{c}_l \rangle \cdot \langle \mathbf{c}_k \rangle} + \sum_{k \neq i} \overline{\langle \mathbf{c}_k^2 \rangle} + 2 \overline{\mathbf{c}_{is} \cdot \sum_{k \neq i} \langle \mathbf{c}_k \rangle} + \overline{\mathbf{c}_{is}^2} \right).$$

So, after some similar algebra for $u_{is}^*(p) \equiv u_i^*(p_{-i}; s)$, we obtain the following comparison between two strategies $s_1, s_2 \in \mathcal{S}$:

$$u^*(p_{-i}; s_2) - u^*(p_{-i}; s_1) = 2 \left[ u_{is_2}^*(p) - u_{is_1}^*(p) \right] + \frac{1}{N} \left( \overline{\mathbf{c}_{is_2}^2} - \overline{\mathbf{c}_{is_1}^2} \right) \quad (12)$$

Now, given the preprogramming (8) of $\mathbf{c}$, we note that $(\mathbf{c}_{is}^m)^2$ takes on the value $\mathbf{q}_r^2 = -1 + \frac{1}{y_r}$ with probability $y_r$. Hence, the central limit theorem (recall that $M = \lambda N = O(N)$) implies that $\frac{1}{M} \sum_{m=1}^M (\mathbf{c}_{is}^m)^2$ will have mean $\sum_r y_r (\frac{1}{y_r} - 1) = B - 1$ and variance $\frac{1}{M} \sum_r (\frac{1}{y_r} - B)$, the latter being negligible unless $\mathbf{y}$ is too close to the faces of $\Delta_B$. More concretely:

*Definition 6:* A distribution $\mathbf{y} \in \text{Int}(\Delta_B)$ is *proper* when $\frac{1}{B-1} \sum_{r=1}^B \left( \frac{1}{y_r} - B \right) = o(1)$; otherwise, $\mathbf{y}$ is called *degenerate*.

Henceforward, our working assumption will be that there are no degenerate nodes: otherwise, we could simply remove them from the analysis (i.e. reduce $B$ and modify $\mathbf{y}$ accordingly). This reflects the fact that degeneracy in the strength distribution simply indicates that certain nodes have extremely low strength scores and all reasonable users shun them.[9]

With this in mind, the last term of (12) will be on average 0 and with a variance of lesser order than the first term. Thus:

$$u^*(p_{-i}; s_2) - u^*(p_{-i}; s_1) \sim 2 \left[ u_i^*(p_{-i}; s_2) - u_i^*(p_{-i}; s_1) \right] \quad (13)$$

i.e. the aggregate payoff $u^*$ is indeed a *potential function* for the game $\mathfrak{G}^*$ (at least asymptotically). We have thus proven:

*Lemma 7:* Let $\mathfrak{G}$ be a simplex game for $N$ players. Then, as $N \to \infty$, the maxima of the averaged aggregate $u^* = \sum_{i=1}^N u_i^*$ will correspond (almost surely) to correlated equilibria of $\mathfrak{G}$.

### C. Anarchy and Efficiency

Still, one expects quite the gulf between anarchic and efficient states: after all, selfish players are hardly the ones to rely upon for social efficiency. In the context of networks, this contrast is frequently measured by the *price of anarchy*, a notion first introduced in [23] as the (coordination) ratio between the maximum attainable aggregate payoff and the one attained at the game's equilibria. Then, depending on whether we look at worst or best-case equilibria, we get the *pessimistic* or *optimistic* price of anarchy respectively.

In our game, the aggregate payoff is equal to: $u = \sum_{i=1}^N u_i = -\frac{1}{N} \sum_{i=1}^N \mathbf{b}_i \sum_{j=1}^N \mathbf{b}_j = -\frac{1}{N} \mathbf{b}^2$ and attains a maximum of $u_{\max} = 0$ when $\mathbf{b} = 0$. So, if we recall by (7) that a Nash equilibrium occurs if and only if $\mathbf{b} = 0$, we see that *Nash anarchy does not impair efficiency*. Clearly, neither the users, nor the agencies that deploy the wireless network could hope for a better solution!

However, this also shows that the traditional definition of the price of anarchy is no longer suitable for our purposes. One reason is that $u_{\max} = 0$ and, hence, we cannot hope to get any information from ratios involving $u_{\max}$.[10] What's more, the users' selfishness in our setting is more aptly captured by the Aumann equilibria of definition 4, so we should be taking the signal-averaged $u^*$ instead of $u$. As a result, we are led to:

*Definition 8:* Let $\mathfrak{G}$ be a simplex game for $N$ players and $B$ nodes. Then, if $p = (p_1 \ldots p_N)$ is a mixed strategy profile of $\mathfrak{G}$, we define its *frustration level* to be:

$$R(p) = -\frac{1}{B-1} u^*(p) = \frac{1}{N(B-1)} \frac{1}{M} \sum_m \mathbf{b}^2(p) \qquad (14)$$

that is, the (average) distance from the Nash solution $\mathbf{b} = 0$. Also, the game's *correlated price of anarchy* $R(\mathfrak{G})$ will be:

$$R(\mathfrak{G}) = \inf \left\{ R(p) : p \in \Delta^0(\mathfrak{G}) \right\} \qquad (15)$$

i.e. the minimum value of the frustration level over the set $\Delta^0(\mathfrak{G})$ of the game's constrained correlated equilibria.

Some remarks are now in order: first and foremost, we see that the frustration level of a strategy profile measures

---

[9]After all, degenerate nodes cannot serve more than $o(\sqrt{N})$ users.

[10]This actually highlights a general problem with the coordination ratio: it does not behave well w.r.t. adding a constant to the payoff functions.

(in)efficiency by contrasting the average aggregate payoff to the optimal case $u_{\max} = 0$ (the normalization $\frac{1}{B-1}$ has been introduced for future convenience). So, with correlated equilibria representing the anarchic states of the game, we remain justified in the eyes of [23] by calling $R(\mathfrak{G})$ the price of anarchy. In effect, the only thing that sets us apart is that, instead of a ratio, we are taking the difference.

Finally, one might wonder why we do not consider the *pessimistic* version by replacing the inf of the above definition with a sup. The main reason for this is that in the next section, we will present a scheme with which users will be able to converge to their *most efficient* equilibrium. Thus, there is no reason to consider worst-case equilibria as in [23]: we only need to measure the price of *sophisticated* anarchy.

## IV. Evolution and Equilibria

Naturally, as the simplex game is iterated, one may assume that rational users will want to maximize their payoff by employing more often the strategies that perform better. The most obvious way to accomplish this is to keep track of a strategy's performance and reward it accordingly:

*Definition 9:* Let $\mathfrak{G}$ be a simplex game as in definition 3, and let $\omega = (m, s_1 \ldots s_N)$ be an instance of $\mathfrak{G}$. Then, the *reward* to the $s^{\text{th}}$ strategy of player $i$ is the random variable:

$$W_{is}(\omega) = \tfrac{1}{M} u_i(m, s_{-i}; s) = -\tfrac{1}{MN} \mathbf{c}_{is}^m \cdot \left[ \mathbf{b}(\omega) + \left( \mathbf{c}_{is}^m - \mathbf{c}_{is_i^m} \right) \right] \quad (16)$$

In other words, the reward $W_{is}$ that player $i$ awards to his $s^{th}$ strategy is (a fraction of) the payoff that the strategy would have garnered for the player in the given instance.[11]

A seeming problem with the above definition is that, in order to learn and evolve, users will have to rate *all* their strategies, i.e. they must be able to calculate the payoff even of strategies they did not employ. So, given that the payoff is a function of the aggregate bet $\mathbf{b}$, it would seem that users would have to be informed of every other user's bet, a prospect that downright shatters the unregulated premises of our setting. However, a more careful consideration of (6) reveals that it suffices for users to know the distribution of users among the nodes, something which is small enough to be broadcasted by the nodes along with the signal $m$.[12]

So, let us consider a sequence $\omega(t)$ of instances of $\mathfrak{G}$ to model the game's $t^{\text{th}}$ iteration ($t = 0, 1, 2 \ldots$). At time $t + 1$, players rank their strategies according to their *scores*:

$$U_{is}(t+1) = U_{is}(t) + W_{is}(\omega(t)) \quad (17)$$

where we set $U_{is}(0) = 0$ to reflect that there is no a priori predisposition towards any given strategy. Then, strategies are selected according to their scores, following the evolutionary scheme of *exponential learning* (see e.g. [18], [27])

$$p_{is}(t) = \frac{e^{\Gamma_i U_{is}(t)}}{\sum_{s'} e^{\Gamma_i U_{is'}(t)}} \quad (18)$$

where $\Gamma_i$ represents the *learning rate* of player $i$.

[11]The rescaling factor $\frac{1}{M}$ has been introduced because significant rewards should come only after checking a strategy against at least $O(M)$ signals.

[12]Actually, the signal itself could be the user distribution of the previous stage. This was discussed in [20] where the distinction between real and fake memory is seen to have a negligible impact on the game's performance.

As a first step to understand the dynamical system of (18), we note that players' evolution actually takes place over the time scale $\tau = t/M$: it takes an average of $O(M)$ iterations to notice a distinct change in the scores of (18). In this case, the score of a strategy will have been modified by: $\delta U_{is} = \sum_{t=\tau}^{\tau+M} W_{is}(\omega(t)) = -\frac{1}{MN} \sum_{t=\tau}^{\tau+M} \left( \mathbf{c}_{is}^{m(t)} \cdot \sum_{j \neq i} \mathbf{c}_{js_j(t)}^{m(t)} + \left( \mathbf{c}_{is}^{m(t)} \right)^2 \right)$. But, by applying the central limit theorem, we may write $\sum_{j \neq i} \mathbf{c}_{js_j(t)}^{m(t)} \sim \sum_{j \neq i} \langle \mathbf{c}_i^{m(t)} \rangle$ and, under some mild ergodicity assumptions, we can also approximate the time average $\frac{1}{M} \sum_{t=\tau}^{\tau+M} (\cdot)$ by the ensemble average $\frac{1}{M} \sum_m (\cdot)$. Thus, the change in a strategy's score after $M$ iterations will be:

$$\delta U_{is} \sim -\tfrac{1}{N} \left[ \mathbf{c}_{is} \cdot \sum_{j \neq i} \langle \mathbf{c}_j \rangle + \mathbf{c}_{is}^2 \right] = u_i^*(p_{-i}; s) \quad (19)$$

A fine point in the above is the implicit assumption that $p_{is}$ changes very slowly. This caveat collapses if the learning rates $\Gamma_i$ are too high (i.e. when we approach "hard" best-response schemes)[13] but, if we stay away from this limit, we may pass to continuous time and differentiate (18) to obtain:

$$\frac{\mathrm{d}p_{is}}{\mathrm{d}\tau} = \Gamma_i p_{is} \left( u_i^*(p) - u_i^*(p_{-i}; s) \right) \quad (20)$$

since, by (19), $\frac{\mathrm{d}U_{is}}{\mathrm{d}\tau}$ will be given by $u_i^*(p_{-i}; s)$.

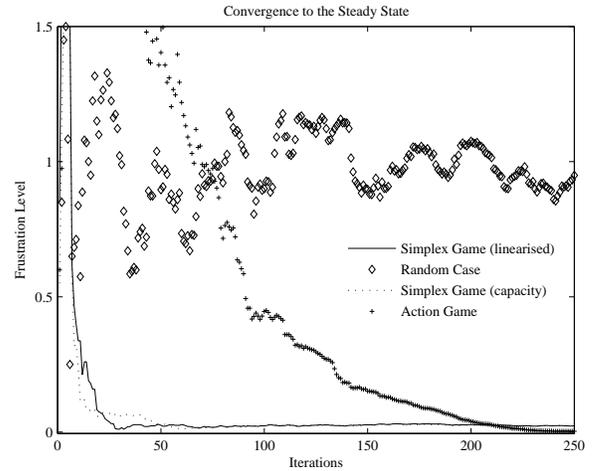

Fig. 1. Simulation of a simplex game for $N = 50$ players that seek to connect to $B = 5$ nodes of random strengths with the help of $M = 2$ broadcasts and $S = 2$ strategies. The game is iterated based on (18) with a learning rate of $\Gamma_i = 20$, and we plot the users' (instantaneous) frustration $R_t = -\frac{u}{B-1}$ (cf. (14)) versus the number of iterations $t$: as predicted by theorem 11, players quickly converge to a steady state of minimal frustration. To justify the linearisation of (4), we also simulated a game with the nonlinear payoff (1), obtaining virtually indistinguishable results. As a baseline, we consider unsophisticated users who simply pick a node randomly, thus experiencing much higher frustration (on average $R = 1$). Finally, we also simulate replicator dynamics (with the same learning rate) on the congestion game determined by (1); in that case, although users eventually reach the Nash solution, they do so at a much slower rate.

These dynamics are extremely powerful: they are the standard multi-population *replicator dynamics* for the correlated form $\mathfrak{G}^*$ of the game. To be sure, in Weibull's extremely comprehensive account [29], it is shown that they exhibit a striking equivalence: the asymptotically stable states[14] of (20)

[13]See [28] for a detailed discussion on this.

[14]These are attracting steady states that are also Lyapunov stable.

are precisely the (strict) Nash equilibria of the underlying game (in our case $\mathfrak{G}^*$). But, since a strategy profile is a Nash equilibrium for the correlated game $\mathfrak{G}^*$ if and only if it is a correlated equilibrium for the original game $\mathfrak{G}$, this proves:

*Lemma 10:* Let $\mathfrak{G}$ be a simplex game, iterated under (18). Then, almost surely as $N \to \infty$, a profile $p = (p_1 \ldots p_N)$ will be asymptotically stable w.r.t. the dynamics of (18) if and only if it is a constrained correlated equilibrium of $\mathfrak{G}$.

So, what remains to be seen is whether the learning scheme of (18) really does lead the game to such a fortuitous state. To that end, one would expect that, as users evolve, they learn how to minimize their average frustration level and eventually settle down to a stable local minimum. Roughly speaking, this is the content of a *Lyapunov function*, i.e. a function $L \equiv L(p)$ with $\frac{dL}{d\tau} \leq 0$. If such a function exists, Lyapunov's theorem will ensure convergence to the steady state and, thankfully, there is an obvious candidate: the aggregate payoff $u^*$ which is also the potential of the correlated game $\mathfrak{G}^*$.

Indeed, if we combine (13) and (20), we can see that: $\frac{du^*}{d\tau} = \sum_i \sum_s \frac{\partial u^*}{\partial p_{is}} \frac{dp_{is}}{d\tau} = \frac{1}{2} \sum_i \Gamma_i \sum_s u^*(p_{-i}; s) \, p_{is} \, (u^*(p_{-i}; s) - u^*(p)) \geq 0$, the last step owing to Jensen's inequality (recall that $u^*(p) = \sum_s p_{is} u^*(p_{-i}; s)$). In other words, the frustration $R = -\frac{1}{B-1} u^*$ is a Lyapunov function for the dynamics of (20) and the players will converge to its global minimum; in effect, this proves:

*Theorem 11:* If a simplex game $\mathfrak{G}$ with a large number $N$ of players is iterated under the exponential learning scheme (18), the players' mixed strategies will converge almost surely to an asymptotically stable state $p^*$ with the following properties:

(i) $p^*$ is a (strict) constrained correlated equilibrium of $\mathfrak{G}$;
(ii) $p^*$ is the most efficient equilibrium of $\mathfrak{G}$, in the sense that it maximizes the aggregate payoff $u^*$ over all $p \in \prod_{i=1}^{N} \Delta_S$;
(iii) $p^*$ is pure.

*Proof:* Thanks to the preceding discussion and Lyapunov's second theorem, we only need to prove part (iii). But, since $u^*$ is harmonic in $p$, it will attain its maximum value on one of the vertices of $\mathcal{D} = \prod_{1}^{N} \Delta_S$. Then, seeing as $p^*$ maximizes $u^*$ by part (ii), it must be pure. ∎

## V. The Price of Anarchy

So far, we have seen that the dynamics of exponential learning lead the users to an evolutionarily stable equilibrium which maximizes (on average) their aggregate payoff (given their preprogramming). Hence, as far as measuring anarchy is concerned, we only need to calculate the level of frustration at this steady state: rather surprisingly, it will turn out that the price of anarchy is *independent* of the distribution $\mathbf{y}$ of the nodes' strengths. In fact, the analytic expression that we obtain at the end of this section shows that it is a function only of the number $B$ of nodes in the network, the training parameter $\lambda = \frac{M}{N}$ and the number $S$ of strategies per user.

To begin with, equation (11) for the frustration level $R$ at a mixed strategy profile $p$ can be rewritten as:

$$R(p) = \frac{1}{N(B-1)} \left[ \sum_i \langle \mathbf{c}_i^2 \rangle + \sum_{i,j \atop i \neq j} \langle \mathbf{c}_i \rangle \cdot \langle \mathbf{c}_j \rangle \right] \quad (21)$$

So, recalling definition 6 and the discussion for the aggregate payoff (12), the first term of (21) will be: $\frac{1}{N(B-1)} \sum_i \langle \mathbf{c}_i^2 \rangle \sim 1$.

Then, to deal with the second term in (21), note that for a given $m$, the aggregate bet $\mathbf{b}(m, p) = \sum_i \langle \mathbf{c}_i^m \rangle$ gives: $\mathbf{b}(m, p)^2 = \sum_i \sum_s p_{is}^2 \left( \mathbf{c}_{is}^m \right)^2 + \sum_i \sum_{s \neq s'} p_{is} p_{is'} \mathbf{c}_{is}^m \cdot \mathbf{c}_{is'}^m + \sum_{i,j \atop i \neq j} \langle \mathbf{c}_i^m \rangle \cdot \langle \mathbf{c}_j^m \rangle$. Thus, to leading order in $N$, this expression has an average of:[15]

$$\frac{1}{M} \sum_m \mathbf{b}(m, p)^2 \sim \sum_{i,j \atop i \neq j} \langle \mathbf{c}_i \rangle \cdot \langle \mathbf{c}_j \rangle + (B-1) \sum_i \sum_s p_{is}^2 \quad (22)$$

As a result, equations (21) and (22) may be combined to:

$$R(p) \sim 1 + \frac{1}{MN(B-1)} \sum_{m=1}^{M} \mathbf{b}(m, p)^2 - G(p) \quad (23)$$

where $G(p) = \frac{1}{N} \sum_i \sum_s p_{is}^2$.

By definition 8, the game's (optimistic) price of anarchy $R(\mathfrak{G})$ will simply be the minimum of $R(p)$ over the game's equilibria. But, since the minimum of $R$ is an equilibrium by theorem 11, we can simply take the minimum over *all* mixed profiles: $R(\mathfrak{G}) = \min\{R(p) : p \in \prod_{i=1}^{N} \Delta_S\}$. In this way, we get a minimization problem of the kind commonly encountered in statistical physics where one seeks to harvest the ground states of (similar in form) energy functionals [30].

Motivated by this, we introduce the *partition function*:

$$\mathscr{Z}(\beta, \mathbf{c}) = \int_{\mathcal{D}} e^{-\beta N R(p)} \, dp \quad (24)$$

where $\mathcal{D} = \prod_{1}^{N} \Delta_S$ and $dp = \prod_{i,s} dp_{is}$ is Lebesgue measure on $\mathcal{D}$.[16] In this way, we may integrate asymptotically to write:[17]

$$R(\mathfrak{G}) = -\frac{1}{N} \lim_{\beta \to \infty} \frac{1}{\beta} \log \mathscr{Z}(\beta, \mathbf{c}). \quad (25)$$

To proceed, we will make the mild (but important) assumption that, for large $N$, it matters little which specific strategy matrix the users actually picked. More formally:

*Assumption 12 (Self-averaging):* For any strategy matrix $\mathbf{c}$:

$$\log \mathscr{Z}(\beta, \mathbf{c}) \sim \langle \log \mathscr{Z}(\beta) \rangle_{\text{all } \mathbf{c}} \quad (26)$$

almost surely as $N \to \infty$ (the averaging $\langle \cdot \rangle$ takes place over all $B^{NSM}$ matrices $\mathbf{c}$, drawn according to (8)).

This is a fundamental assumption in statistical physics and describes the rarity of configurations which yield notable differences in macroscopically observable parameters. Under this light, we are left to calculate $\langle \log \mathscr{Z} \rangle$, a problem which we will attack with the help of *replica analysis*.[18]

The starting point of the method is the identity $\langle \log \mathscr{Z} \rangle = \lim_{a \to 0^+} \frac{1}{a} \log \langle \mathscr{Z}^a \rangle$ which reduces the problem to powers of $\mathscr{Z}$:[19]

$$R(\mathfrak{G}) = -\frac{1}{N} \lim_{\beta \to \infty} \lim_{a \to 0^+} \frac{1}{a\beta} \langle \mathscr{Z}^a(\beta) \rangle \quad (27)$$

These are much more manageable since, for $n \in \mathbb{N}$:

$$\mathscr{Z}^n = \left( \int_{\mathcal{D}} e^{-\beta N R(p)} \, dp \right)^n = \int_{\mathcal{D}} \cdots \int_{\mathcal{D}} e^{-\beta N \sum_\mu R(p_\mu)} \prod_\mu dp_\mu \quad (28)$$

---

[15] See also [21] (pp. 529) for more on this point.

[16] $\mathscr{Z}$ depends on the strategy matrix $\mathbf{c}$ through the frustration level $R(p)$.

[17] Essentially, this refers to the fact that $\max_D f = \lim_{s \to \infty} \frac{1}{s} \log \int_D e^{sf(t)} \, dt$ for any measurable function $f$ on a compact domain $D$ (see e.g. [31]).

[18] See [30] for a general discussion or [21], [32] for the minority game.

[19] To prove this identity, write $\mathscr{Z}^a = e^{a \log \mathscr{Z}}$ and expand.

i.e. $\mathscr{Z}^n = \prod_{\mu=1}^n \mathscr{Z}_\mu$, where $\mathscr{Z}_\mu = \int_{\mathscr{D}} \exp\left(-N\beta R(p_\mu)\right) \mathrm{d}p_\mu$ is the partition function for the $\mu^{\text{th}}$ *replica* $p_\mu = \{p_{is\mu}\}$ of the system. Then, thanks to equation (23), we obtain:

$$\langle \mathscr{Z}^n(\beta) \rangle = A^n \int_{\mathscr{D}^n} \left\langle e^{-\frac{\beta}{M(B-1)} \sum_\mu \sum_m (\mathbf{c}_\mu^m)^2} \right\rangle e^{N\beta \sum_\mu G_{\mu\nu}(p)} \prod_\mu \mathrm{d}p_\mu \quad (29)$$

where $A = e^{-N\beta}$; $\mathbf{c}_\mu^m = \mathbf{b}(m, p_\mu) = \sum_i \sum_s p_{is\mu} \mathbf{c}_{is}^m$ is the aggregate bet for the mixed profile $p_\mu = \{p_{is\mu}\}$ in the $\mu^{\text{th}}$ replica (given the signal $m$); and $G_{\mu\nu}(p) = \frac{1}{N} \sum_i \sum_s p_{is\mu} p_{is\nu}$.

Of course, what we really need is to express $\langle \mathscr{Z}^n \rangle$ for real values of $n$ in the vicinity of $n = 0^+$; for this, we resort to:

*Assumption 13 (Replica Continuity):* The expression given in (29) for $\langle \mathscr{Z}^n \rangle$ can be continued analytically to all real values of $n$ in the vicinity of $n = 0^+$.

At first glance, this might appear as a blind leap of faith, especially since uniqueness criteria (e.g. log-convexity) are absent. However, such criteria can in some cases be established (see e.g. [33]) and, moreover, the huge amount of literature surrounding this assumption and the agreement of our own analysis with our numerical results (see figures 2–5) makes us feel justified in employing it.

With the help of the above, and after the lengthy calculations of appendix B, we are in a position to prove:

*Theorem 14 (Irrelevance of Node Strengths):* Let $\mathbf{y}, \mathbf{y}' \in \text{Int}(\Delta_B)$ be strength distributions for $B$ nodes and let $\mathfrak{G}, \mathfrak{G}'$ be simplex games for $\mathbf{y}$ and $\mathbf{y}'$ respectively. Then, as $N \to \infty$:

$$R(\mathfrak{G}) \sim R(\mathfrak{G}') \quad (30)$$

In other words, we are (rather unexpectedly!) reduced to the symmetric case of $B$ equivalent nodes: ceteris paribus, *the price of anarchy depends only on the number of nodes present and not on their individual strengths*.

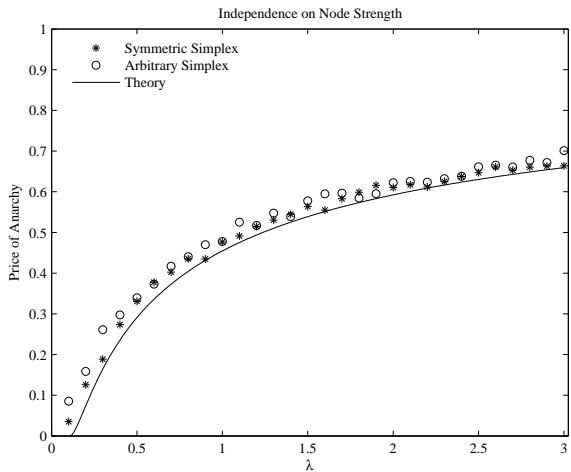

Fig. 2. The price of anarchy (i.e. the steady-state frustration level) as a function of the training parameter $\lambda = \frac{M}{N}$ for $B = 4$ equivalent nodes contrasted to that of 4 nodes employing standards with different spectral efficiencies $c_r$: EVDO-Rev.A (1.06 Mbps), HSDPA (3.91 Mbps), 802.11b (11Mbps) and WiMAX (14.1 Mbps) [34]; we simulated $N = 50$ users with $S = 2$ strategies and averaged over 25 realizations of the game. As predicted by theorem 14, different standards do not affect the price of anarchy.

Now, in order to actually determine the *effect of choices* on the users' frustration level, we first define the *binary reduction*

of a simplex game $\mathfrak{G}$ for $B$ nodes. This is just a simplex game $\mathfrak{G}_{\text{eff}}$ for 2 identical nodes and a training set enlarged by $B-1$, i.e. $M_{\text{eff}} = M(B-1)$; everything else remains the same. Then, under this rescaling, the same train of calculations that is used to prove theorem 14 also yields:

*Theorem 15 (Reduction of Choices):* The price of anarchy for a simplex game $\mathfrak{G}$ is asymptotically equal to that of its binary reduction $\mathfrak{G}_{\text{eff}}$; in other words, as $N \to \infty$:

$$R(\mathfrak{G}) \sim R(\mathfrak{G}_{\text{eff}}) \quad (31)$$

Thanks to this equivalence, we see that the price of anarchy depends on $M$ and $B$ only through $M(B-1)$; so, for example, if some nodes go offline, we will know exactly how much to increase $M$ so as to maintain the same performance level.

However, theorem 15 really tells us much more: it provides a "dictionary" between the simplex game and the extensively studied minority game. Indeed, mutatis mutandis, one sees that the price of anarchy $R(\mathfrak{G})$ corresponds to the *market volatility* $\sigma$ in the minority game [21]. So, if we follow the *(replica-symmetric)* calculations of [21], we finally obtain the price of anarchy in terms of the game's parameters $B$, $S$ and $\lambda = \frac{M}{N}$:

$$R(\mathfrak{G}) \sim \Theta(\lambda - \lambda_c)\left(1 - \sqrt{\lambda_c/\lambda}\right)^2 \quad (32)$$

where $\Theta$ is the Heaviside step function and $\lambda_c = \lambda_c(S, B)$ is the critical value that marks the emergence of anarchy within the premises of replica symmetry (see appendix B).[20]

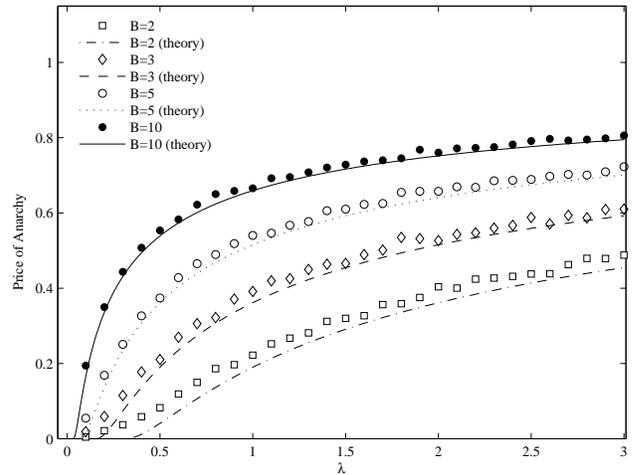

Fig. 3. The effect of choices: we plot the price of anarchy as a function of the training parameter $\lambda = M/N$ for $N = 50$ users, $S = 2$ strategies and $B = 2, 3, 5, 10$ nodes (averaging over 25 realizations). We see that efficiency *deteriorates* as $B$ increases: more choices actually confuse the users.

This expression is one of our key results since it accurately captures the impact of the various system parameters on the network's performance (see e.g. figures 2–5). So, even though it follows effortlessly by virtue of theorem 15, for the sake of completeness (and also to discuss the role of *replica symmetry*), we carry out the derivation of (32) in appendix B.

---

[20]Actually $\lambda_c = \frac{\zeta^2(S)}{B-1}$ with $\zeta(S) = S/2^{S-1}\sqrt{2/\pi} \int_{-\infty}^{\infty} z e^{-z^2} \text{erfc}^{S-1}(z) \, \mathrm{d}z$.

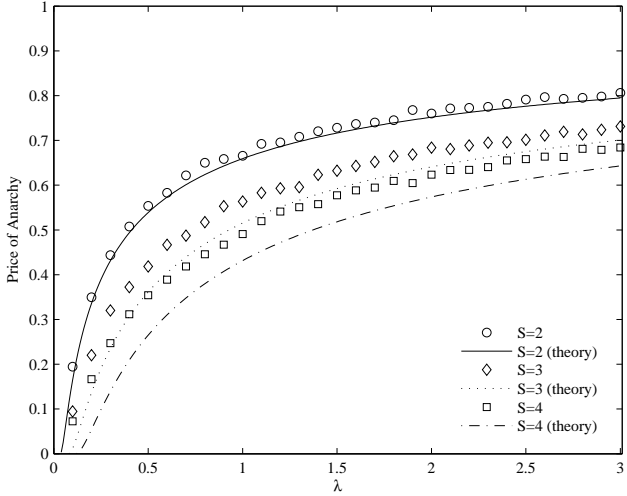

Fig. 4. The effect of sophistication: we plot the price of anarchy as a function of the training parameter $\lambda = M/N$ for $N = 50$ users, $S = 2, 3, 4$ strategies and $B = 5$ nodes (again averaging over 25 realizations of the game). As expected, sophisticated users (larger $S$) are more efficient.

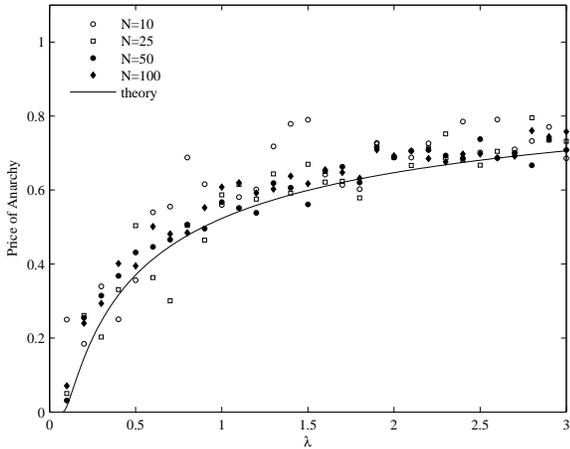

Fig. 5. The price of anarchy as a function of the training parameter $\lambda = M/N$ for different numbers of users $N = 10, 25, 50, 100$, with $B = 5$ nodes and $S = 2$ strategies; unlike other plots, we are harvesting the price of anarchy from a *single* realization of the game. We see that the number of players does not seriously impact the price of anarchy (except through $\lambda$).

## VI. Conclusions

Our main goal was to analyze an unregulated network of (a large number of) heterogeneous users that can connect to a multitude of wireless nodes with different specifications (e.g. different standards). In such a network, users who selfishly try to maximize their individual downlink throughput (2) will have to compete against each other for the nodes' finite resources. So, in the pursuit of order (and in the absence of a central overseer), we advocate the use of a training beacon (such as a random integer synchronously broadcasted by the nodes) to act as a coordination stimulus: by processing this stimulus with the aid of some preprogrammed strategies and choosing a node accordingly, users should be able to reach an equilibrium.

Indeed, if users keep records of their strategies' performance and rank them based on the evolutionary scheme of exponential learning (18), they learn to coordinate their actions and quickly reach an evolutionarily stable state.[21] This state is also *socially* stable in the sense that unilateral deviation is (on average) discouraged: it is a *correlated equilibrium*. Then, to measure the efficiency of users in this setting, we examine how far they are from the optimal distribution that maximizes their aggregate throughput. In so doing, we see that *exponential learning leads the users to their most efficient equilibrium*.

However, since the users' rationality is bounded (i.e. they can only handle a small number of strategies), this equilibrium will still be at some distance from the optimal state. This distance is the *price of (correlated) anarchy* and we calculate it with the method of replicas. Interestingly, we find (theorem 14) that *the price of anarchy does not depend on the nodes' characteristics, but only on their number*. In fact, we provide a reduction of our scenario to the minority game [19] (theorem 15) and, as a result, we obtain the analytic expression (32) for the price of anarchy. This also generalizes the results obtained for the minority game to an arbitrary number of choices.

Thanks to the above, we derive quantitative predictions about the degree of anarchy in our scenario. For example (fig. 3), we see that blindly adding more nodes to a network is not a panacea: anarchy actually *increases* with the number of nodes because the users are not able to process the added complexity and do not make efficient use of the extra resources. On the other hand, if users become more sophisticated and employ more strategies (fig. 4), anarchy comes at a lesser price (albeit at a slower convergence to a stable state). Finally, we see that the number of users really doesn't have to be quite so large (fig. 5): these conclusions hold even for the much smaller numbers of users typically encountered in local service areas.

## Appendix A
### Properties of y-Simplices

We begin here by showing that definition 1 is not vacuous:

*Lemma 16:* There exists a **y**-simplex $\mathscr{B} = \{\mathbf{q}_r\}_{r=1}^{B} \subseteq \mathbb{R}^{B-1}$ for any $\mathbf{y} \in \mathrm{Int}(\Delta_B)$.

*Proof:* Begin by selecting a vector $\mathbf{q}_1 \in \mathbb{R}^B$ such that $\mathbf{q}_1^2 = \frac{1}{y_1} - 1$ and choose $\mathbf{q}_{r+1} \in \mathbb{R}^B$ inductively so that it satisfies (5) when multiplied by $\mathbf{q}_1 \ldots \mathbf{q}_r$. Such a selection is always possible for $r \leq B - 1$ thanks to the dimension of $\mathbb{R}^B$; for a vector space of lesser dimension, this is no longer the case. [22]

In this way, we obtain $B$ vectors $\mathbf{q}_r \in \mathbb{R}^B$ that satisfy (5); our construction will be complete once we show that $\mathscr{B}$ is contained in some $(B-1)$-subspace of $\mathbb{R}^B$. However, as in the proof of lemma 2, we can see that $\sum_{r=1}^{B} y_r \mathbf{q}_r = 0$; this means that $\mathscr{B}$ is linearly dependent and completes our proof. ∎

The next lemma is a key property of **y**-simplices that plays a crucial role in the calculations of appendix B:

---

[21] In figure 1 we see that convergence occurs within tens of iterations. Thus, if each iteration is of the order of milliseconds (a reasonable transmission timescale for wideband wireless networks), this corresponds to equilibration times of tens of milliseconds.

[22] Note that $\frac{1}{y_r y_l} \geq \frac{1}{y_r} + \frac{1}{y_l}$ for all $r, l$ so that $\mathbf{q}_r^2 \cdot \mathbf{q}_l^2 \geq (\mathbf{q}_r \cdot \mathbf{q}_l)^2$ holds for (5).

*Lemma 17:* Let $\mathscr{B} = \{\mathbf{q}_r\}_{r=1}^{B} \subseteq \mathbb{R}^{B-1}$ be a **y**-simplex for some $\mathbf{y} \in \text{Int}(\Delta_B)$. Then, for all $\mathbf{x} \in \mathbb{R}^{B-1}$: $\sum_{r=1}^{B} y_r (\mathbf{q}_r \cdot \mathbf{x})^2 = \mathbf{x}^2$.

*Proof:* Since $\mathbf{y} \in \text{Int}(\Delta_B)$, $\mathscr{B}$ will span $\mathbb{R}^{B-1}$ and $\mathbf{x}$ may be written as a linear combination $\mathbf{x} = \sum_{r=1}^{B} x_r \mathbf{q}_r$. So, if we let $S = \sum_{r=1}^{B} x_r$ and recall that $\sum_{r=1}^{B} y_r = 1$, we will have: $\mathbf{x}^2 = \sum_{l,r=1}^{B} x_l x_r \mathbf{q}_r \cdot \mathbf{q}_l = -S^2 + \sum_{r=1}^{B} x_r^2 / y_r$. Similarly: $(\mathbf{q}_r \cdot \mathbf{x})^2 = S^2 - 2S \frac{x_r}{y_r} + \frac{x_r^2}{y_r^2}$, and an addition over $r$ yields the lemma. ∎

## Appendix B
### Measuring the Price of Anarchy

Picking up where we left off in section V, we begin by calculating the expression for $\langle \mathscr{Z}^n \rangle$ in (29). To do this, we will use the identity: $e^{-\frac{\mathbf{q}^2}{2}} = \frac{1}{(2\pi)^{k/2}} \int_{\mathbb{R}^k} e^{i\mathbf{q}\cdot\mathbf{z} - \frac{\mathbf{z}^2}{2}} \, d\mathbf{z} = \mathbf{E}_{\mathbf{z}} e^{i\mathbf{q}\cdot\mathbf{z}}$ where $\mathbf{E}_{\mathbf{z}}$ denotes expectation over a Gaussian random vector $\mathbf{z}$ with $k$ independent components $z_1 \ldots z_k \sim N(0,1)$; this is the *Hubbard-Stratonovich* transformation. So, if $\{\mathbf{z}_{\mu}^m = (z_{\mu,1}^m, \ldots z_{\mu,B-1}^m)\}_{\mu=1\ldots n}^{m=1\ldots M}$ are such vectors of $\mathbb{R}^{B-1}$, we get:

$$\left\langle e^{-\frac{\beta}{M(B-1)} \sum_{\mu} \sum_m (\mathbf{c}_{\mu}^m)^2} \right\rangle = \mathbf{E}_{\{\mathbf{z}_{\mu}^m\}} \left\langle e^{i \sum_i \sum_{\mu} \sum_m \mathbf{x}_{is}^m \cdot \mathbf{c}_{is}^m} \right\rangle \quad (33)$$

where: $\mathbf{x}_{is}^m = \sqrt{\frac{2\beta}{M(B-1)}} \sum_{\mu} p_{is\mu} \mathbf{z}_{\mu}^m \in \mathbb{R}^{B-1}$. Then, by the independence of the $\mathbf{c}_i$'s (eq. (8)), we will be able to obtain the average $\langle \cdot \rangle$ of (33) over the matrices $\mathbf{c}$ by computing the characteristic function $\left\langle e^{i\mathbf{x}\cdot\mathbf{q}} \right\rangle$ for only one of them. This is done in the following:

*Lemma 18:* Let $\mathbf{y} \in \text{Int}(\Delta_B)$ and let $\mathscr{B} = \{\mathbf{q}_r\}_{r=1}^{B}$ be a **y**-simplex in $\mathbb{R}^{B-1}$. If $\mathbf{x} \in \mathbb{R}^{B-1}$ and $\mathbf{q}$ is a random vector with distribution $P(\mathbf{q} = \mathbf{q}_r) = y_r$, then: $\left\langle e^{i\mathbf{x}\cdot\mathbf{q}} \right\rangle = e^{-\frac{\mathbf{x}^2}{2}} + O\left(|\mathbf{x}|^3\right)$.

*Proof:* Expanding the exponential $\langle \exp(i\mathbf{x}\cdot\mathbf{q}) \rangle$ yields:

$$\begin{aligned}
\left\langle e^{i\mathbf{x}\cdot\mathbf{q}} \right\rangle &= \left\langle 1 + i\mathbf{x}\cdot\mathbf{q} - \tfrac{1}{2}(\mathbf{x}\cdot\mathbf{q})^2 + O\left(|\mathbf{x}|^3\right) \right\rangle \\
&= 1 + i\mathbf{x}\cdot\sum_r y_r \mathbf{q}_r - \tfrac{1}{2} \sum_r y_r (\mathbf{q}_r \cdot \mathbf{x})^2 + O\left(|\mathbf{x}|^3\right) \\
&= 1 - \tfrac{1}{2}\mathbf{x}^2 + O\left(|\mathbf{x}|^3\right) = e^{-\frac{1}{2}\mathbf{x}^2} + O\left(|\mathbf{x}|^3\right) \quad (34)
\end{aligned}$$

where the third equality comes from lemmas 2 and 17. ∎

In our case, $|\mathbf{x}_{is}^m| = O(M^{-\frac{1}{2}})$; so, if we apply the previous lemma to each of the random vectors $\mathbf{c}_{is}^m$, the average of eq. (33) will become (to leading order in $N$):

$$\left\langle e^{i \sum_i \sum_m \mathbf{x}_{is}^m \cdot \mathbf{c}_{is}^m} \right\rangle \sim e^{-\frac{1}{2} \sum_{i,s,m} (\mathbf{x}_{is}^m)^2} = e^{-\frac{\beta}{M(B-1)} \sum_m \sum_{\mu,\nu} G_{\mu\nu}(p) \, \mathbf{z}_{\mu}^m \cdot \mathbf{z}_{\nu}^m} \quad (35)$$

where $\lambda = \frac{M}{N}$ is the game's training parameter.

Now, if we introduce the $n \times n$ matrix $\mathbf{J} = \mathbf{I} + \frac{2\beta}{\lambda(B-1)} \mathbf{G}(p)$, and recall that $\int_{\mathbb{R}^n} e^{-\frac{1}{2} \sum_{\mu,\nu} J_{\mu\nu} w_{\mu} w_{\nu}} \widetilde{dw} = |\det(\mathbf{J})|^{-\frac{1}{2}}$,[23] we may integrate over the auxiliary variables $\mathbf{z}_{\mu}^m$ to obtain:

$$\begin{aligned}
\mathbf{E}_{\mathbf{z}_{\mu}^m} \left\langle e^{-\frac{\beta}{M(B-1)} \sum_{\mu} \sum_m (\mathbf{c}_{\mu}^m)^2} \right\rangle &\sim \int_{\mathbb{R}^{nM(B-1)}} e^{-\frac{1}{2} \sum_{m=1}^{M} \sum_{k=1}^{B-1} \sum_{\mu,\nu} J_{\mu\nu}(p) \, z_{\mu,k}^m z_{\nu,k}^m} \widetilde{dz} \\
&= \left( \int_{\mathbb{R}^n} e^{-\frac{1}{2} \sum_{\mu,\nu} J_{\mu\nu}(p) \, w_{\mu} w_{\nu}} \widetilde{dw} \right)^{M(B-1)} = e^{-\frac{M(B-1)}{2} \log \det(\mathbf{J}(\mathbf{p}))} \quad (36)
\end{aligned}$$

So, after these calculations, equation (29) finally becomes:

$$\frac{\langle \mathscr{Z}^n(\beta) \rangle}{A^n} \sim \int_{\mathscr{D}^n} e^{N\beta \left[ \text{tr}(\mathbf{G}(p)) - \frac{N(B-1)}{2\beta} \log \det \left( \mathbf{I} + \frac{2\beta}{\lambda(B-1)} \mathbf{G}(p) \right) \right]} \prod_{\mu} dp_{\mu} \quad (37)$$

---

Clearly, this last expression is independent of the strength distribution **y**, a fact which proves theorem 14. In addition, we observe that (37) remains invariant when we pass from the game $\mathfrak{G}$ to its binary reduction $\mathfrak{G}_{\text{eff}}$ with the rescaled training parameter $\lambda_{\text{eff}} = \lambda(B-1)$, thus proving theorem 15 as well.

Now, to proceed from (37), we will introduce $n^2$ $\delta$-functions in their integral representation so as to isolate the profiles $p_{is}$: $\delta\left(\mathbf{Q} - \mathbf{G}(p)\right) = \left(\frac{N\beta}{2\pi}\right)^{n^2} \int e^{iN\beta \sum_{\mu,\nu} k_{\mu\nu}(Q_{\mu\nu} - G_{\mu\nu}(p))} \prod_{\mu,\nu} dk_{\mu\nu}$. In this way, the integral of (37) becomes:

$$\int e^{-N\beta \left[ \frac{\lambda(B-1)}{2\beta} \log \det \left( \mathbf{I} + \frac{2\beta}{\lambda(B-1)} \mathbf{Q} \right) - \text{tr}(\mathbf{Q}) - i \sum_{\mu,\nu} k_{\mu\nu}(Q_{\mu\nu} - G_{\mu\nu}(p)) \right]} d\sigma \quad (38)$$

where $d\sigma = \prod_{\mu} dp_{\mu} \times \prod_{\mu,\nu} dk_{\mu\nu} \times \prod_{\mu,\nu} dQ_{\mu\nu}$ is the product measure on $\mathscr{D}^n \times \mathbb{R}^{n^2} \times \mathbb{R}^{n^2}$. However, $p$ only appears in the last term of the (38) and can be integrated separately to yield:

$$\begin{aligned}
\int_{\mathscr{D}^n} e^{-iN\beta \sum_{\mu,\nu} k_{\mu\nu} G_{\mu\nu}(p)} \prod_{\mu} dp_{\mu} &= \prod_{i=1}^{N} \left( \int_{\Delta_S^n} e^{-i\beta \sum_{\mu,\nu} k_{\mu\nu} \sum_s p_{is\mu} p_{is\nu}} \prod_{s,\mu} dp_{is\mu} \right) \\
&= \exp\left( N \log \int_{\Delta_S^n} e^{-i\beta \sum_{\mu,\nu} k_{\mu\nu} \sum_s p_{s\mu} p_{s\nu}} \prod_{s,\mu} dp_{s\mu} \right) \quad (39)
\end{aligned}$$

(recall that $G_{\mu\nu}(p) = \frac{1}{N} \sum_i \sum_s p_{is\mu} p_{is\nu}$ and $\mathscr{D}^n = (\Delta_S)^{N \times n}$). So, by descending to the limit $N \to \infty$, we find:

$$\begin{aligned}
\frac{1}{N} \log \left\langle \mathscr{Z}^n(\beta) \right\rangle \sim -\beta \Big[ n + \frac{\lambda(B-1)}{2\beta} \log \det \left( \mathbf{I} + \frac{2\beta}{\lambda(B-1)} \mathbf{Q} \right) - \text{tr}(\mathbf{Q}) - \\
- i \sum_{\mu,\nu} k_{\mu\nu} Q_{\mu\nu} - \frac{1}{\beta} \log \int_{\Delta^n} e^{-i\beta \sum_{\mu,\nu} k_{\mu\nu} \sum_s p_{s\mu} p_{s\nu}} \prod_{s,\mu} dp_{s\mu} \Big] =: -\beta \Lambda \quad (40)
\end{aligned}$$

where $\mathbf{Q}$ and $\mathbf{k}$ extremize the function $\Lambda$ within the brackets. This is where we will invoke *replica symmetry* (see [30], [32]).

*Assumption 19 (Replica Symmetry):* The *saddle-points* of $\Lambda$ are of the form:

$$Q_{\mu\nu} = q + (Q - q)\,\delta_{\mu\nu}; \quad k_{\mu\nu} = i\lambda\beta\frac{B-1}{2}\left(r + (R - r)\,\delta_{\mu\nu}\right) \quad (41)$$

In other words, we seek saddle-point matrices that are symmetric in the replica space (the scaling factors are there for future convenience).[24] Under this ansatz, we obtain:

$$\begin{aligned}
\Lambda = n + \frac{\lambda(B-1)}{2\beta} \log \det \left( \frac{2\beta}{\lambda} \frac{q}{B-1} + \left( 1 + \frac{2\beta}{\lambda} \frac{Q-q}{B-1} \right) \delta_{\mu\nu} \right) - nQ \\
+ n\lambda\beta\frac{B-1}{2}(QR - qr) + n^2\lambda\beta\frac{B-1}{2}qr \\
- \frac{1}{\beta} \log \int_{\Delta^n} e^{\lambda\beta^2 \frac{B-1}{2} \left((R-r) \sum_{\mu} \mathbf{p}_{\mu}^2 + r \left(\sum_{\mu} \mathbf{p}_{\mu}\right)^2\right)} \prod_{s,\mu} dp_{s\mu} \quad (42)
\end{aligned}$$

where $\mathbf{p}_{\mu}$ is the generic profile $(p_{1\mu} \ldots p_{S\mu})$ in the $\mu^{\text{th}}$ replica.

The second term of the above expression can be easily calculated by noting that $\det\left(q + p\delta_{\mu\nu}\right) = p^n \left(1 + n\frac{q}{p}\right)$: it will be equal to $\frac{q}{1+\chi} + \frac{\lambda}{2\beta}(B-1)\log(1+\chi) + o(n)$, where $\chi = \frac{2\beta}{\lambda} \frac{Q-q}{B-1}$. As for the last term of (42), we will again use the Hubbard-Stratonovich transformation with a canonical Gaussian variable $\mathbf{z}$ of $\mathbb{R}^S$ to write: $e^{r\lambda\beta^2 \frac{B-1}{2} \left(\sum_{\mu} \mathbf{p}_{\mu}\right)^2} = \mathbf{E}_{\mathbf{z}} e^{\sqrt{r\lambda(B-1)} \mathbf{z} \cdot \sum_{\mu} \mathbf{p}_{\mu}}$. Then, for notational convenience, we also let:

$$V(\mathbf{z}, \mathbf{p}) = \sqrt{r\lambda(B-1)}\, \mathbf{z} \cdot \mathbf{p} - \lambda\beta\frac{B-1}{2}(R - r)\mathbf{p}^2 \quad (43)$$

---

[23] Here, tildes as in $\widetilde{dw}$ denote Lebesgue measure normalized by $\sqrt{2\pi}$.

[24] This assumption can actually be dropped; e.g. see [32] where the first step of *symmetry breaking* (1RSB) is performed. Still, replica symmetry does not incur a significant error on our calculations while greatly simplifying them.

and, in this way, the integral of (42) becomes $(\mathrm{d}p = \prod_1^S \mathrm{d}p_s)$:

$$\log \mathbf{E_z} \int_{\Delta_S^n} e^{-\beta \sum_\mu V(\mathbf{z}, \mathbf{p}_\mu)} \prod_{s,\mu} \mathrm{d}p_{s\mu} = n \, \mathbf{E_z} \log \int_{\Delta_S} e^{-\beta V(\mathbf{z}, \mathbf{p})} \mathrm{d}p + o(n)$$

From (27) and the premises of replica continuity (assumption 13), what we really need to calculate is $\Lambda_0 = \lim_{n\to 0} \frac{1}{n}\Lambda$:

$$\Lambda_0 = 1 + \frac{q}{1+\chi} + \frac{\lambda}{2\beta}(B-1)\log(1+\chi) - Q$$
$$+ \lambda\beta\frac{B-1}{2}(QR - qr) - \frac{1}{\beta}\,\mathbf{E_z}\left[\log\int_{\Delta_S} e^{-\beta V(\mathbf{z},\mathbf{p})}\,\mathrm{d}p\right] \quad (44)$$

where $Q, q, R, r$ have been chosen so as to satisfy the *replica-symmetric* saddle-point equations: $\frac{\partial\Lambda_0}{\partial Q} = 0$, $\frac{\partial\Lambda_0}{\partial q} = 0$, etc.

To that end, it can be shown that both $Q - q$ and $R - r$ are of order $O(1/\beta)$, i.e. $\chi$ remains finite as $\beta \to \infty$. So, in this limit, we will once again perform asymptotic integration for the integrals of $\frac{\partial\Lambda_0}{\partial R} = 0$ and $\frac{\partial\Lambda_0}{\partial r} = 0$. Thus, we are led to consider the vertex $\mathbf{p}_*(\mathbf{z})$ of $\Delta_S$ which minimizes the harmonic function $V(\mathbf{z}, \cdot)$ and we obtain:

$$Q \sim \phi \qquad\qquad R = r + \frac{1}{\beta}\frac{2}{\lambda(B-1)}\frac{\chi}{1+\chi} \quad (45)$$
$$q \sim \phi + \frac{1}{\beta}\frac{\zeta}{\sqrt{\lambda(B-1)}r} \qquad r = \frac{4}{\lambda^2(B-1)^2}\frac{1}{(1+\chi)^2}$$

where $\phi = \mathbf{E_z}[\mathbf{p}_*^2(\mathbf{z})]$ and $\zeta = \mathbf{E_z}[\mathbf{p}_*(\mathbf{z})\cdot\mathbf{z}]$.

Now, if we let $\beta \to \infty$ and substitute (45) in (44), we get $\Lambda_0 = 1 - \phi + \phi\left(1 + \zeta/\sqrt{\phi\lambda(B-1)}\right)^2$ where, after a little geometry: $\zeta(S) = \mathbf{E_z}[\min\{z_1 \ldots z_S\}] = \frac{S}{2^{S-1}}\sqrt{\frac{2}{\pi}}\int_{-\infty}^\infty z e^{-z^2}\mathrm{erfc}^{S-1}(z)\,\mathrm{d}z$ and $\phi = 1$. Hence, for finite $\chi$ (i.e. for $\lambda \geq \lambda_c = \frac{\zeta^2(S)}{B-1}$), we finally acquire expression (32) for the game's *price of anarchy*:

$$R(\mathfrak{G}) \sim \Lambda_0 \sim \Theta(\lambda - \lambda_c)\left(1 - \sqrt{\lambda_c/\lambda}\right)^2.$$